\documentclass[10pt]{article}
\usepackage{amssymb}
\usepackage{amsmath}
\usepackage{amscd}
\usepackage{dashrule}
\usepackage[svgnames,table]{xcolor}
\usepackage{booktabs}
\usepackage{misccorr}
\usepackage{indentfirst}
\def\be{\begin{equation}}
\def\ee{\end{equation}}
\def\arr{\begin{array}{rll}}
\def\ea{\end{array}}
\def\bea{\begin{eqnarray}}
\def\eea{\end{eqnarray}}

\def\N2{$N{=}2$}

\def\>{\rangle}
\def\<{\langle}
\def\+{\dagger}
\def\={\ =\ }

\begin{document}
\renewcommand{\thefootnote}{\arabic{footnote}}
\begin{center}
    {\LARGE\bf Towards $\ell$-conformal Galilei algebra via}
    \vskip 0.3cm 
    {\LARGE\bf contraction of the conformal group}
    \vskip 1cm
    $
    \textrm{\Large Ivan Masterov}
    $
    \vskip 0.3cm
    {\it
    Tomsk Polytechnic University,\\
    634050, Tomsk, Lenin Ave. 30, Russia}
    \vskip 0.5cm
    {E-mail: masterov@tpu.ru}
\end{center}

\vskip 0.3cm
\begin{center}
    $\sim\circ\sim\bullet\sim\circ\sim$ -------------------------------------------------------------- $\sim\circ\sim\bullet\sim\circ\sim$
    
    {\bf Abstract}
\end{center}

\noindent
\leftskip=1.5cm
\rightskip=1.5cm
We show that the In\"{o}n\"{u}-Wigner contraction of $so(\ell+1,\ell+d)$ with the integer $\ell>1$ may lead to algebra which contains a variety of conformal extensions of the Galilei algebra as subalgebras. These extensions involve the $\ell$-conformal Galilei algebra in $d$ spatial dimensions as well as $l$-conformal Galilei algebras in one spatial dimension with $l=3$, $5$, ..., $(2\ell-1)$.
\vskip 0.3cm


\begin{center}
    --------------------------------------------------------------------------------------------------------------------
\end{center}

\leftskip=0cm
\rightskip=0cm
\vskip 0.3cm
\section{Introduction}

Seventy years ago, in 1953, E. In\"{o}n\"{u} and E. P. Wigner published the now classical work \cite{Wigner} in which the method for contracting Lie algebras was introduced. The method is based on a singular transformation which does not change a number of generators and preserves the structure relations of some chosen subalgebra. All generators, which are out from this subalgebra, begin to form an abelian ideal.

The In\"{o}n\"{u}-Wigner ($\mathbb{IW}$) contraction provides a way to show how some Lie algebras can be obtained as a limiting cases of others. This is a useful tool which allows us to realize the nonrelativistic limit of relativistic algebras \cite{Wigner,Bacry}. For example, the nonrelativistic limit of the (anti) de Sitter algebra is the so-called Newton-Hooke algebra. The $\mathbb{IW}$ contraction of the Poincar\'{e} algebra yields the Galilei one.

The presence of an abelian ideal in a nonrelativistic algebra may motivate a search for a possible relativistic counterpart of this algebra. However, the existence of a solution to this problem is not guaranteed. In 1972, U. Niederer established that the maximal kinematical invariance group of the free Schr\"{o}dinger equation is the Schr\"{o}dinger group which, besides the Galilei transformations, contains dilatations and special conformal transformations  \cite{Niederer_1972}. This result provoked a series of works in which authors investigated possible relationships between the Schr\"{o}dinger algebra and the relativistic conformal one \cite{Barut}-\cite{Havas}. Despite all efforts, a relativistic counterpart in the sense of an $\mathbb{IW}$ contraction for the Schr\"{o}dinger algebra has, to date, not been found, and probably does not exist.

Trying to get the Schr\"{o}dinger symmetry from the relativistic conformal one, the new conformal extension of the Galilei algebra was derived as the nonrelativistic limit of $so(2,4)$ \cite{Barut,Havas} (see also \cite{Lukierski}). In the literature this extension is referred to as the conformal Galilei algebra. By comparison with the Schr\"{o}dinger algebra, the conformal Galilei one involves the extra vector generator associated with accelerations.

Realizations of both conformal extensions in a nonrelativistic spacetime are characterized by the following representation of the dilatations' generator:
\bea
&&
L_{0} = t\frac{\partial}{\partial t} + \frac{1}{z} x_i \frac{\partial}{\partial x_i},
\nonumber
\eea
where the constant $z$ is the rational dynamical exponent which takes the value $1$ for the conformal Galilei algebra and value $2$ for the Schr\"{o}dinger algebra. In 1997, M. Henkel \cite{Henkel} and J. Negro et al. \cite{Negro_1,Negro_2} established that any nonzero value of $z$ may be associated with its own conformal extension of the Galilei algebra. But generally, such an extension is infinite-dimensional. The requirement of finite dimensionality is satisfied if, and only if, the parameter $\ell=1/z$ is a positive integer or half-integer. The corresponding finite-dimensional extension is called the $\ell$-conformal Galilei algebra. The first two options $\ell=1/2$ and $\ell=1$ correspond to the Schr\"{o}dinger algebra and the conformal Galilei one respectively. 

In recent years the instances with $\ell>1$ have been intensively studied in the context of their various realizations. In particular, applications in condensed matter physics have been discussed in \cite{Henkel_2002}-\cite{Henkel_2004}. Higher derivative mechanics with such symmetries has been considered in the works \cite{DH}-\cite{EA}. Dynamical systems without higher derivative terms have been constructed in \cite{AV_2013,AV_2012} (see also \cite{FIL}). Spacetimes with the $\ell$-conformal Galilei symmetry have been explored in \cite{AV_2016}-\cite{CS}.  Recent progress has been related with conformally invariant perfect fluid dynamics \cite{AV_2022}-\cite{Snegirev} (see also \cite{HZ}).

The $\ell$-conformal Galilei algebra involves the abelian ideal which consists of $(2\ell+1)$ vector generators. So, the necessary condition for the possibility to obtain this algebra from another by contraction is fulfilled for all $\ell$. However, at the moment a relativistic counterpart in the sense of an $\mathbb{IW}$ contraction is known only for $\ell=1$.

An interesting observation was made in the work \cite{Negro_2} by analysing the matrix representation of the $\ell$-conformal Galilei algebra in three spatial dimensions.  It was noticed that for integer $\ell$ this representation can be obtained from a contraction of the conformal algebra $so(\ell+1,\ell+3)$. However, an explicit form of such contraction procedure was not presented. Note that the number of generators in the $\ell$-conformal Galilei algebra for $\ell>1$ is less than the dimension of its supposed relativistic partner. Therefore, one may expect that the $l$-conformal Galilei algebra for $\ell>1$ may arise only as a subalgebra of a contracted algebra. If this is true, the natural question arises: what structure does a contracted algebra have? The purpose of the present work is to formulate the contraction procedure for $so(\ell+1,\ell+d)$ with the integer $\ell>1$ which leads the $\ell$-conformal Galilei algebra in $d$ spatial dimensions as a subalgebra.

The paper is organized as follows. In Sect. \ref{sect2} we recall some basic facts about the $\ell$-conformal Galilei algebra and the conformal algebra $so(\ell+1,\ell+d)$. In Sect. \ref{sect3} we obtain an appropriate basis of $so(\ell+1,\ell+d)$. In Sect. \ref{sect4} we formulate an $\mathbb{IW}$ contraction procedure for this algebra with respect to $so(1,2)\oplus so(d)$ subalgebra. The case $\ell=2$ is treated in more detail. In particular, we show how a realization of $so(3,2+d)$ in a flat spacetime can be contracted. We summarize our results and discuss further possible developments in the concluding Sect. \ref{sect5}.

\section{The $\ell$-conformal Galilei algebra and $so(\ell+1,\ell+d)$}\label{sect2}

In this section we recall some basic facts about the $\ell$-conformal Galilei algebra and the conformal algebra $so(\ell+1,\ell+d)$.

\vskip 0.2cm

The $\ell$-conformal Galilei algebra in $d$ spatial dimensions contains \cite{Henkel,Negro_1}
\bea
&&
\begin{aligned}
    &
    \circ \mbox{ the generator of time translations} && L_{-1};
    \\[2pt]
    &
    \circ \mbox{ the generator of dilatations} && L_{0};
    \\[2pt]
    &
    \circ \mbox{ the generator of special conformal transformations} && L_{1};
    \\[2pt]
    &
    \circ \mbox{ the chain of vector generators} && C_i^{(n)},\quad n = \overline{0,2\ell},\; i = \overline{1,d};
    \\[2pt]
    &
    \circ \mbox{ the generators of spatial rotations} && M_{ij},\quad i,j = \overline{1,d}.
\end{aligned}
\nonumber
\eea
The generators $C_i^{(0)}$ and $C_i^{(1)}$ correspond to spatial translations and Galilei boosts while other vector generators are linked to accelerations. The non-vanishing structure relations in the algebra read
\bea\label{Galilei}
&&
\begin{aligned}
    &
    [L_{n},L_{m}] = (m-n)L_{n+m}, && [L_{n},C_i^{(m)}] = (m-\ell(n+1)) C_i^{(n+m)},
    \\[2pt]
    &
    [M_{ij},C_{k}^{(n)}] = \delta_{ik} C_j^{(n)} - \delta_{jk} C_{i}^{(n)}, && [M_{ij},M_{kp}] = -\delta_{ik} M_{jp} + \delta_{ip} M_{jk} + \delta_{jk} M_{ip} - \delta_{jp} M_{ik}.
\end{aligned}
\eea
Note that $L_{-1}$, $C_i^{(0)}$, $C_{i}^{(1)}$ and $M_{ij}$ form the Galilei algebra. The generators $L_{-1}$, $L_{0}$, $L_{1}$ obey the commutation relations of the conformal algebra in one dimension $so(1,2)$.

The $\ell$-conformal Galilei algebra \eqref{Galilei} can be realized in a flat $(d+1)$-dimensional spacetime in terms of the operators:
\bea\label{Galilei_gen}
&&
L_{n} = t^{n+1}\frac{\partial}{\partial t} + \ell (n+1) x_i \frac{\partial}{\partial x_i}, \qquad C_i^{(n)} = t^{n} \frac{\partial}{\partial x_i}, \qquad M_{ij} = x_i\frac{\partial}{\partial x_j} - x_j \frac{\partial}{\partial x_i}.
\eea

Let us consider a flat spacetime with the metric\footnote{Here and in what follows we consider only integer values of  $\ell$ which are more than one.}
\bea
&&
g_{\alpha\beta} = \mbox{diag}(\underbrace{-,-,...,-}_{\mbox{$\ell$ times}},\underbrace{+,+,...,+}_{\mbox{ $\ell+d-1$ times}}).
\nonumber
\eea
As is known, conformal transformations of this spacetime involve such transformations as preserve the metric up to a local scale factor. Their generators read
\bea\label{so_gen}
&&
P_{\alpha}=\frac{\partial}{\partial y^{\alpha}}, \quad D = y^{\alpha}\frac{\partial}{\partial y^{\alpha}},\quad
K_{\alpha} = y^{\beta} y_{\beta} \frac{\partial}{\partial y^{\alpha}}-2y_{\alpha} y^{\beta}\frac{\partial}{\partial y^{\beta}},\quad
J_{\alpha\beta} = y_{\beta}\frac{\partial}{\partial y^{\alpha}} - y_{\alpha}\frac{\partial}{\partial y^{\beta}}.
\eea
These generators obey the following commutation relations
\bea\label{kommut}
&&
\begin{aligned}
&
[D,P_{\alpha}]=-P_{\alpha}, \quad [D,K_{\alpha}]=K_{\alpha}, \quad [P_{\alpha},K_{\beta}]=-2(J_{\alpha\beta}+\eta_{\alpha\beta}D),
\\[2pt]
&
[J_{\alpha\beta},P_{\gamma}]=\eta_{\alpha\gamma}P_{\beta}-\eta_{\beta\gamma}P_{\alpha}, \quad [J_{\alpha\beta},K_{\gamma}]=\eta_{\alpha\gamma}K_{\beta}-\eta_{\beta\gamma}K_{\alpha},
\\[2pt]
&
[J_{\alpha\beta},J_{\mu\nu}] = \eta_{\alpha\mu} J_{\beta\nu} - \eta_{\alpha\nu} J_{\beta\mu} - \eta_{\beta\mu} J_{\alpha\nu} + \eta_{\beta\nu} J_{\alpha\mu}.
\end{aligned}
\eea

It can be shown that the algebra \eqref{kommut} is isomorphic to $so(\ell+1,\ell+d)$ (see, e.g., \cite{Qualls}) whose structure relations have the form
\bea\label{so}
[\mathbb{M}_{\alpha\beta},\mathbb{M}_{\mu\nu}] = \eta_{\alpha\mu} \mathbb{M}_{\beta\nu} - \eta_{\alpha\nu} \mathbb{M}_{\beta\mu} - \eta_{\beta\mu} \mathbb{M}_{\alpha\nu} + \eta_{\beta\nu} \mathbb{M}_{\alpha\mu},
\eea
where
\bea\label{eta}
&&
\eta_{\alpha\beta}= \mbox{ diag} (\underbrace{-,-,..,-}_{\mbox{$\ell+1$ times}},\underbrace{+,+,..,+}_{\mbox{$\ell+d$ times}})
\eea
and all indices take the values $\overline{0,2\ell+d}$.

\section{An appropriate basis of $so(\ell+1,\ell+d)$}\label{sect3}

In this section we obtain such basis of $so(\ell+1,\ell+d)$ for the integer $\ell>1$ in which, after applying a contraction procedure, some subsets of generators will obey the structure relations \eqref{Galilei}. For simplicity, we designate $so(\ell+1,\ell+d)$-analogues of the Galilean generators by the same letters as in \eqref{Galilei}.

\subsection{$M_{ij}$, $L_0$ and a grading of the subalgebra $so(\ell+1,\ell)$}\label{subs3.1}

In this subsection we introduce $so(\ell+1,\ell+d)$-analogues for the Galilean generators $M_{ij}$ and $L_0$. Apart from that, we obtain the grading for the subalgebra $so(\ell+1,\ell)$, which in particular will be helpful in the next subsection where we will derive counterparts for the generators $L_{-1}$ and $L_1$ .

\vskip 0.2cm

By comparing \eqref{Galilei} and \eqref{so} we may immediately put
\bea\label{M}
&&
M_{ij} = -\mathbb{M}_{2\ell+i,2\ell+j}.
\eea
Taking into account that the generator $L_{0}$ commutes with $M_{ij}$, it is natural to seek an analogue of this scalar among elements of the subalgebra $so(\ell+1,\ell)$ spanned by $\mathbb{M}_{\alpha\beta}$ with $\alpha,\beta=\overline{0,2\ell}$. Let us introduce
\bea\label{L0}
&&
L_{0} = \sum_{\alpha=0}^{\ell-1}(\ell-\alpha)\mathbb{M}_{\alpha,2\ell-\alpha}
\eea
and find eigenvectors under the adjoint action of $L_0$ on $so(\ell+1,\ell)$. In other words, we are interested in such elements as $A\in so(\ell+1,\ell)$ which satisfy the relation
\bea
&&
ad_{L_0}A \equiv [L_0,A] = \lambda A,
\nonumber
\eea
where an eigenvalue $\lambda$ is a constant which we will call {\bf the weight} of $A$. 

Each pair $(\mathbb{M}_{\alpha \ell},\mathbb{M}_{\ell,2\ell-\alpha})$ as well as each quartet of generators $(\mathbb{M}_{\alpha\beta},\mathbb{M}_{\beta,2\ell-\alpha},\mathbb{M}_{\alpha,2\ell-\beta},\mathbb{M}_{2\ell-\beta,2\ell-\alpha})$ with fixed $\alpha,\beta = \overline{0,\ell-1}$ are closed under the adjoint action of $L_0$:
\bea
\begin{aligned}
&
[L_0,\mathbb{M}_{\alpha \ell}] = (\ell-\alpha)\mathbb{M}_{\ell,2\ell-\alpha}, && [L_0,\mathbb{M}_{\ell,2\ell-\alpha}] = (\ell-\alpha)\mathbb{M}_{\alpha\ell},
\\[2pt]
&
[L_0,\mathbb{M}_{\alpha\beta}] = (\ell-\alpha)\mathbb{M}_{\beta,2\ell-\alpha} - (\ell-\beta)\mathbb{M}_{\alpha,2\ell-\beta}, && [L_0,\mathbb{M}_{\beta,2\ell-\alpha}] = (\ell-\alpha)\mathbb{M}_{\alpha\beta} - (\ell-\beta)\mathbb{M}_{2\ell-\beta,2\ell-\alpha},
\\[2pt]
&
[L_0,\mathbb{M}_{2\ell-\beta,2\ell-\alpha}] = (\ell-\alpha)\mathbb{M}_{\alpha,2\ell-\beta} - (\ell-\beta)\mathbb{M}_{\beta,2\ell-\alpha}, && [L_0,\mathbb{M}_{\alpha,2\ell-\beta}] = (\ell-\alpha)\mathbb{M}_{2\ell-\beta,2\ell-\alpha} - (\ell-\beta)\mathbb{M}_{\alpha\beta}.
\end{aligned}
\nonumber
\eea
As a consequence, it can easily be shown that
\bea\label{wedges}
&&
[L_0,\Omega_{\alpha}^{\pm}] = \pm (\ell-\alpha) \Omega_{\alpha}^{\pm}, \qquad [L_0,\Lambda_{\alpha\beta}] = (\beta-\alpha)\Lambda_{\alpha\beta}, \qquad [L_0,\Sigma_{\alpha\beta}^{\pm}] = \pm (2\ell-\alpha-\beta)\Sigma_{\alpha\beta}^{\pm},
\eea
where we denote
\bea\label{basis}
&&
\begin{aligned}
&
\Omega_{\alpha}^{\pm} = \mathbb{M}_{\alpha \ell}\pm \mathbb{M}_{\ell,2\ell-\alpha}, 
\\[2pt]
&
\Lambda_{\alpha\beta} = \mathbb{M}_{\alpha\beta}+\mathbb{M}_{2\ell-\beta,2\ell-\alpha} + \mathbb{M}_{\beta,2\ell-\alpha}+\mathbb{M}_{\alpha,2\ell-\beta},
\\[2pt]
&
\Sigma_{\alpha\beta}^{\pm} = (\mathbb{M}_{\alpha\beta}-\mathbb{M}_{2\ell-\beta,2\ell-\alpha}) \pm (\mathbb{M}_{\beta,2\ell-\alpha}-\mathbb{M}_{\alpha,2\ell-\beta}).
\end{aligned}
\eea
The nonvanishing commutation relations between these generators are given by
\bea\label{comrel}
&&
\begin{aligned}
&
\begin{aligned}
&
[\Lambda_{\alpha \beta},\Lambda_{\mu\nu}] = 2\delta_{\beta\mu}\Lambda_{\alpha\nu} - 2\delta_{\alpha\nu}\Lambda_{\mu\beta}, && \quad [\Lambda_{\alpha\beta},\Omega_{\gamma}^{\pm}] = 2\delta_{\beta\gamma}\Omega_{\alpha}^{\pm},
\\[2pt]
&
[\Sigma_{\alpha\beta}^{\pm},\Omega_{\gamma}^{\mp}] = 2\delta_{\beta\gamma} \Omega_{\alpha}^{\pm} -2\delta_{\alpha\gamma} \Omega_{\beta}^{\pm}, && \quad [\Omega_{\alpha}^{+},\Omega_{\beta}^{-}] = -\Lambda_{\alpha\beta},
\\[2pt]
&
[\Lambda_{\alpha\beta},\Sigma_{\mu\nu}^{\pm}] = 2\delta_{\beta\mu} \Sigma_{\alpha\nu}^{\pm} - 2\delta_{\beta\nu} \Sigma_{\alpha\mu}^{\pm}, && \quad [\Omega_{\alpha}^{\pm},\Omega_{\beta}^{\pm}] = -\Sigma_{\alpha\beta}^{\pm},
\end{aligned}
\\[2pt]
&
[\Sigma_{\alpha \beta}^{+},\Sigma_{\mu\nu}^{-}] = -2\delta_{\alpha\mu} \Lambda_{\beta\nu} + 2\delta_{\alpha\nu} \Lambda_{\beta\mu} + 2\delta_{\beta\mu} \Lambda_{\alpha\nu} - 2\delta_{\beta\nu} \Lambda_{\alpha\mu}.
\end{aligned}
\eea
Note that $\Sigma_{\beta\alpha}^{\pm} = -\Sigma_{\alpha\beta}^{\pm}$ while $\Lambda_{\beta\alpha} \neq \pm \Lambda_{\alpha\beta}$. In terms of $\Lambda_{\alpha\beta}$ the expression \eqref{L0} for the generator $L_0$ can be rewritten as
\bea
&&
L_0 = \frac{1}{2}\sum_{\alpha=0}^{\ell-1}(\ell-\alpha)\Lambda_{\alpha\alpha}.
\eea

The generators $\Omega_{\alpha}^{\pm}$, $\Lambda_{\alpha\beta}$ and $\Sigma_{\alpha\beta}^{\pm}$ form a new basis in the subalgebra $so(\ell+1,\ell)$. Let us denote a subspace which is spanned by generators with the weight $\lambda$ as $V_{\lambda}$. To illustrate this notion, for $\ell=2$, $3$, and $4$ we collect in Table 1 all possible options for $\lambda$ and corresponding basis elements of $V_{\lambda}$.
\begin{table}[ht]\label{table}
  \caption{Basis elements of $V_{\lambda}$ for $\ell=2$, $3$, and $4$}  
  \centering
  \setlength{\extrarowheight}{2pt}  
  \rowcolors{2}{green!10}{white}
  \begin{tabular}{*{1}{>{$}c<{$}}|*{2}{>{$}l<{$}}|*{3}{>{$}l<{$}}|*{4}{>{$}l<{$}}}\toprule
     \lambda &  \multicolumn{2}{c}{$\ell=2$} & \multicolumn{3}{c}{$\ell=3$} & \multicolumn{4}{c}{$\ell=4$} \\\midrule
     -7 & &  & & &  & \Sigma_{01}^{-} & & & \\[2pt]
     -6 & &  & & &  & \Sigma_{02}^{-} & & & \\[2pt]
     -5 & &  & \Sigma_{01}^{-} & &  & \Sigma_{03}^{-}, & \Sigma_{12}^{-} & & \\[2pt]
     -4 & &  & \Sigma_{02}^{-} & &  & \Sigma_{13}^{-}, & \Omega_{0}^{-} & & \\[2pt]
     -3 & \Sigma_{01}^{-} &  & \Sigma_{12}^{-}, & \Omega_{0}^{-} &  & \Sigma_{23}^{-}, & \Omega_{1}^{-}, & \Lambda_{30} & \\[2pt]
     -2 & \Omega_{0}^{-} &  & \Omega_{1}^{-}, & \Lambda_{20} &  & \Omega_{2}^{-}, & \Lambda_{20}, & \Lambda_{31} & \\[2pt]
     -1 & \Omega_{1}^{-}, & \Lambda_{10}  & \Omega_{2}^{-}, & \Lambda_{10}, & \Lambda_{21}  & \Omega_{3}^{-}, & \Lambda_{10}, & \Lambda_{21}, & \Lambda_{32} \\[2pt]
     0 &  \Lambda_{00}, & \Lambda_{11}  & \Lambda_{00}, & \Lambda_{11}, & \Lambda_{22}  & \Lambda_{00}, & \Lambda_{11}, & \Lambda_{22}, & \Lambda_{33} \\[2pt]
     1 &  \Omega_{1}^{+}, & \Lambda_{01}  & \Omega_{2}^{+}, & \Lambda_{01}, & \Lambda_{12}  & \Omega_3^{+}, & \Lambda_{01}, & \Lambda_{12}, & \Lambda_{23} \\[2pt]
     2 & \Omega_{0}^{+} &  & \Omega_{1}^{-}, & \Lambda_{02} &  & \Omega_2^{+}, & \Lambda_{02}, & \Lambda_{13} & \\[2pt]
     3 & \Sigma_{01}^{+} &  & \Sigma_{12}^{+}, & \Omega_{0}^{+} &  & \Sigma_{23}^{+}, & \Omega_{1}^{+}, & \Lambda_{03} & \\[2pt]
     4 &  &  & \Sigma_{02}^{+} & &  & \Sigma_{13}^{+}, & \Omega_{0}^{+} & & \\[2pt]
     5 &  &  & \Sigma_{01}^{+} & &  & \Sigma_{03}^{+}, & \Sigma_{12}^{+} & & \\[2pt]
     6 &  &  & & &  & \Sigma_{02}^{+} & & & \\[2pt]
     7 &  &  & & &  & \Sigma_{01}^{+} & & & \\\bottomrule
  \end{tabular}
\end{table}

In general, for given $\ell$ the weight of generators \eqref{basis} takes the values $0$, $\pm 1$, $\pm2$, ..., $\pm(2\ell-1)$. Consequently, we have the following decomposition:
\bea\label{decom}
so(\ell+1,\ell) = V_{-2\ell+1} \oplus V_{-2\ell+2} \oplus ... \oplus V_{2\ell-1}, \quad \mbox{dim}(V_{\pm(2\ell-2k+1)}) = \mbox{dim}(V_{\pm(2\ell-2k)}) = k, \; k = \overline{1,\ell}.
\eea
Moreover, with the aid of the Jacoby identity for the triple $(L_{0},A_{1},A_{2})$ where $A_{i}\in V_{\lambda_i}$, it is easy to show that $[A_1,A_2]\in V_{\lambda_1+\lambda_2}$. So, the decomposition \eqref{decom} is the grading of the subalgebra $so(\ell+1,\ell)$.

\subsection{$L_{-1}$, $L_{1}$ and an appropriate basis in the subalgebra $so(\ell+1,\ell)$}

In this subsection we obtain $so(\ell+1,\ell+d)$-analogues of the generators $L_{-1}$ and $L_{1}$ whose adjoint action allows us to reveal the $l$-conformal Galilean structures with $l=3,5,...,(2\ell-1)$ in the subalgebra $so(\ell+1,\ell)$.

\vskip 0.2cm

According to \eqref{Galilei}, the generators $L_{-1}$ and $L_{1}$ commute with $M_{ij}$ and have the weights $-1$ and $1$ respectively. Therefore, we conclude that $L_{\pm 1}\in V_{\pm1}$. By analysing the relations \eqref{wedges}, we establish that
\bea
&&
V_{-1} = \mbox{span}(\Lambda_{10},\Lambda_{21},...,\Lambda_{\ell-1,\ell-2},\Omega_{\ell-1}^{-}), \qquad V_{1} = \mbox{span}(\Lambda_{01},\Lambda_{12},...,\Lambda_{\ell-2,\ell-1},\Omega_{\ell-1}^{+}).
\nonumber
\eea
As a consequence, the most general ansatz for $L_{-1}$ and $L_{1}$ is given by 
\bea\label{Lpm1}
&&
L_{-1} = \sum_{\alpha=0}^{\ell-2} \eta_{\alpha}^{-}\Lambda_{\alpha+1,\alpha} + \eta_{\ell-1}^{-}\Omega_{\ell-1}^{-}, \qquad L_{1} = \sum_{\alpha=0}^{\ell-2}\eta_{\alpha}^{+}\Lambda_{\alpha,\alpha+1} + \eta_{\ell-1}^{+}\Omega_{\ell-1}^{+},
\eea
where $\eta_\alpha^{\pm}$ with $\alpha=\overline{0,\ell-1}$ are constants. Then we require that the commutation relation $[L_{-1},L_{1}] = 2L_{0}$ holds. As a result, we obtain the following system of equations:
\bea\label{sys}
&&
\left\{
\begin{aligned}
    &
    \eta_0^{-} \eta_0^{+}  && = -\ell/2,
    \\[2pt]
    &
    \eta_{\alpha}^{-} \eta_{\alpha}^{+} - \eta_{\alpha-1}^{-}\eta_{\alpha-1}^{+} && = (\alpha-\ell)/2, && \alpha = \overline{1,\ell-2},
    \\[2pt]
    &
    \eta_{\ell-1}^{-} \eta_{\ell-1}^{+} + 2\eta_{\ell-2}^{-}\eta_{\ell-2}^{+} && = 0.
\end{aligned}
\right.
\eea
From this it follows that
\bea\label{sol}
&&
\eta_{\alpha}^{-}\eta_{\alpha}^{+} = \left\{
\begin{aligned}
    &
    -\frac{(2\ell-\alpha)(\alpha+1)}{4}, && \alpha = \overline{0,\ell-2},
    \\[2pt]
    &
    \frac{\ell(\ell+1)}{2}, && \alpha = \ell-1.
\end{aligned}
\right.
\eea
So, the system \eqref{sys} has infinitely many solutions. We can fix all constants $\eta_\alpha^{-}$ by any nonzero values while the coefficients $\eta_\alpha^{+}$ can be found with the aid of \eqref{sol}. 

By construction, the generators $L_{0}$ in \eqref{L0} and $L_{\pm1}$ in \eqref{Lpm1} obey the structure relations of $so(1,2)$. Moreover, taking into account that the decomposition \eqref{decom} is the grading of $so(\ell+1,\ell)$, the generators $L_{-1}$ and $L_{1}$ can be associated with the weight lowering operator and the weight raising operator, respectively. It means that for every $A\in V_{\lambda}$ the relation $[L_{\pm1},A] \in V_{\lambda\pm 1}$ is performed.

To obtain a more appropriate basis in the subalgebra $so(\ell+1,\ell)$, let us discuss eigenvectors for $ad_{L_{-1}}$ which correspond to zero eigenvalue. For a given $\ell$, there are existing $\ell$ linear independent null eigenvectors whose weights are distinct and take the values $-1$, $-3$, ..., $-(2\ell-1)$. The instances with the weights $-1$ and $-(2\ell-1)$ are proportional to $L_{-1}$ and $\Sigma_{01}^{-}$ respectively. These are the only null eigenvectors for $\ell=2$. Let us show how we may obtain others for $\ell>2$.

Let us denote the null eigenvector with the weight $\lambda = -(2\ell-2k-1)$ as $T_{2\ell-2k-1}^{(0)}$. 
The most general ansatz for $T_{2\ell-2k-1}^{(0)}$ with $k=0,\,1,...,\,\left[\ell/2\right] - 1$ reads
\bea\label{T1}
&&
T_{2\ell-2k-1}^{(0)} = \sum_{\gamma=0}^{k} a_{\gamma} \Sigma_{\gamma,2k-\gamma+1}^{-},
\eea
where $a_0$, $a_1$, ..., $a_{k}$ are constants, $[\ell/2]$ is an integer part of $\ell/2$.
For $k=0$, the condition $[L_{-1},T_{2\ell-2k+1}^{(0)}] = 0$ is satisfied for arbitrary $a_0$, while for other values of $k$ we derive the homogeneous system of $k$ linear equations in $(k+1)$ variables:
\bea\label{sys1}
&&
a_\gamma\eta_{2k-\gamma}^{-}+a_{\gamma+1}\eta_{\gamma}^{-} = 0, \qquad \gamma = \overline{0,k-1}.
\eea
The coefficient matrix of this system has the rank $k$ and therefore the system has infinitely many nontrivial solutions with one free variable.

The ansatz for $T_{2\ell-2k-1}^{(0)}$ with $k=[\ell/2],\, [\ell/2]+1,\,...,\,\ell-2$ can be written as follows
\bea\label{T2}
&&
T_{2\ell-2k-1}^{(0)} = a_{0}\Omega_{2k-\ell+1}^{-} + \sum_{\gamma=0}^{\ell-k-2}a_{\gamma+1} \Sigma_{2k-\ell+\gamma+2,\ell-\gamma-1}^{-} + \sum_{\gamma=0}^{2k-\ell}a_{\ell-k+\gamma}\Lambda_{2\ell-2k+\gamma-1,\gamma},
\eea
The formal symbol $\sum\limits_{\gamma=0}^{-1}f(\gamma)$, which appears in the latter sum for odd $\ell$ and $k=[\ell/2]$, is assumed to be equal to zero. Let us require that the condition $[L_{-1},T_{2\ell-2k+1}^{(0)}] = 0$ be satisfied. And again we obtain a homogeneous system of $k$ linear independent equations in $(k+1)$ variables
\bea\label{sys2}
&&
\left\{
\begin{aligned}
    &
    a_0 \eta_{2k-\ell}^{-} - a_k \eta_{\ell-1}^{-} && = 0,
    \\[2pt]
    &
    a_0 \eta_{\ell-1}^{-} - 2 a_1 \eta_{2k-\ell+1}^{-} && = 0,
    \\[2pt]
    &
    a_{\gamma} \eta_{\ell-\gamma-1}^{-} + a_{\gamma+1}\eta_{2k-\ell+\gamma+1}^{-} && = 0, && \gamma = \overline{1,\ell-k-2},
    \\[2pt]
    &
    a_{\ell-k+\gamma} \eta_{2\ell-2k+\gamma-1}^{-} - a_{\ell-k+\gamma+1} \eta_{\gamma}^{-} && = 0, && \gamma = \overline{0,2k-\ell-1},
\end{aligned}
\right.
\eea
where the first equation occurred only if $2k-\ell\geq 0$. So, this system also has infinitely many solutions with one free variable. As an illustration, let us write down the null eigenvectors which arise from the solutions of the systems \eqref{sys1} and \eqref{sys2} for $\ell = 3$ and $4$:
\begin{table}[ht]
  \centering
  \setlength{\extrarowheight}{2pt}
  \begin{tabular}{*{1}{>{$}c<{$}}*{1}{>{$}c<{$}}*{1}{>{$}c<{$}}*{1}{>{$}c<{$}}}
      & \ell=3 & \qquad\quad & \ell=4 \\[7pt]
      & T_{5}^{(0)} = a_0\Sigma_{01}^{-}, & & T_{7}^{(0)} = a_0 \Sigma_{01}^{-}, \\[5pt]
      & \displaystyle T_{3}^{(0)} = a_0 \left(\Omega_0^{-} + \frac{\eta_2^{-}}{2\eta_0^{-}} \Sigma_{12}^{-}\right), & & \displaystyle T_{5}^{(0)} = a_0\left(\Sigma_{03}^{-} - \frac{\eta_2^{-}}{\eta_{0}^{-}}\Sigma_{12}^{-}\right), \\[2pt]
      & & & \displaystyle T_{3}^{(0)} = a_0\left(\Omega_{1}^{-}+\frac{\eta_3^{-}}{2\eta_1^{-}}\Sigma_{23}^{-} + \frac{\eta_{0}^{-}}{\eta_3^{-}}\Lambda_{30}\right).
  \end{tabular}
\end{table}

At the next step, let us introduce the following chain of generators
\bea
&&
T_{l}^{(n)} \equiv (-1)^{n}\frac{(2l-n)!}{(2l)!}\underbrace{[L_{1},[L_{1},...[L_{1},T_{l}^{(0)}]...]]}_{\mbox{$n$ times}}, \qquad n=1,2,...,2l
\nonumber
\eea
for every $T_{l}^{(0)}$ with $l =3$, $5$, ..., $(2\ell-1)$. Note that the relation
\bea
&&
[L_{1},T_{l}^{(n)}] = (n-2l) T_{l}^{(n+1)}
\nonumber
\eea
holds by construction. Taking into account that the adjoint action of $L_1$ increases the weight of any element from $so(\ell+1,\ell)$ per unit, the weight of $T_{l}^{(n)}$ is equal to $(n-l)$, i.e.
\bea
&&
[L_{0},T_{l}^{(n)}] = (n-l)T_{l}^{(n)}.
\nonumber
\eea

Let us prove by induction that
\bea\label{rel}
&&
[L_{-1},T_{l}^{(n)}] = n T_{l}^{(n-1)}.
\eea
Firstly,
\bea
&&
[L_{-1},T_{l}^{(1)}] = -\frac{1}{2l}[L_{-1},[L_{1},T_{l}^{(0)}]] = \frac{1}{2l}\left([L_{1},[T_{l}^{(0)},L_{-1}]] + [T_{l}^{(0)},[L_{-1},L_{1}]]\right) = T_{l}^{(0)}.
\nonumber
\eea
Here we use the Jacobi identity for the triple $(L_{-1},L_{1},T_{l}^{(0)})$.

Suppose that the commutator \eqref{rel} holds. Then
\bea
&&
[L_{-1},T_{l}^{(n+1)}] = \frac{1}{n-2l}[L_{-1},[L_{1},T_{l}^{(n)}]] = \frac{1}{2l-n} ([L_{1},[T_{l}^{(n)},L_{-1}]] + [T_{l}^{(n)},[L_{-1},L_{1}]]) = (n+1) T_{l}^{(n)}.
\nonumber
\eea
Thus, the relation \eqref{rel} is proved.

The generators $L_{-1}$, $L_{0}$, $L_{1}$, $T_{l}^{(n)}$ with $l = 3$, $5$, ..., $(2\ell-1)$ and $n=0$, $1$, ..., $2l$ are linear independent. To prove this, it is enough to show that generators with one and the same weight have this property. Let us consider, for example, the generators with the weight $\lambda=-3$ for the case $\ell = 5$: $T_{3}^{(0)}$, $T_{5}^{(2)}$, $T_{7}^{(4)}$, $T_{9}^{(6)}$. Let us suppose that these generators are linear dependent, i.e. that there are existing constants $c_{1}$, $c_2$, $c_3$, and $c_4$, not all zero, such that
\bea
&&
T = c_1 T_{3}^{(0)} + c_2 T_{5}^{(2)} + c_3 T_{7}^{(4)} + c_4 T_{9}^{(6)} \equiv 0.
\nonumber
\eea
But then
\bea
&&
\begin{aligned}
&
0 \equiv \underbrace{[L_{-1},...[L_{-1},T]...]}_{\mbox{6 times}} = 6! c_4 T_{9}^{(0)} && \Rightarrow && c_4 = 0;
&&
0 \equiv \underbrace{[L_{-1},...[L_{-1},T]...]}_{\mbox{4 times}} = 4! c_3 T_{7}^{(0)} && \Rightarrow && c_3 = 0;
\\[2pt]
&
0 \equiv [L_{-1},[L_{-1},T]] = 2! c_2 T_{5}^{(2)} && \Rightarrow && c_2 = 0; && 0\equiv c_1 T_{3}^{(0)} && \Rightarrow && c_1=0.
\end{aligned}
\nonumber
\eea
Therefore, the generators $T_{3}^{(0)}$, $T_{5}^{(2)}$, $T_{7}^{(4)}$, $T_{9}^{(6)}$ are linear independent. The general case can be treated likewise.

Since the number of generators $L_{-1}$, $L_{0}$, $L_{1}$, $T_{l}^{(n)}$ is equal to $\ell(2\ell+1)$ and coincides with the dimension of $so(\ell+1,\ell)$, these generators form the basis of the subalgebra $so(\ell+1,\ell)$. Moreover, the commutation relations between $L_{n}$ with $n= 0,\pm 1$ and others
\bea\label{LT}
&&
[L_{n},T_l^{(m)}] = (m-l(n+1))T_l^{(m+n)}, \quad l = 3,5,...,(2\ell-1),\; m = 0,1,...,2l,
\eea
are the same as in the $l$-conformal Galilei algebras in one spatial dimension. Of course the commutators $[T_{l_1}^{(n_1)},T_{l_2}^{(n_2)}]$ do not vanish here but their explicit form is not interested in the work because the generators $\{T_{l}^{(n)}\}$ will form an abelian ideal after performing a contraction procedure.

\subsection{Analogues of vector generators $C_i^{(n)}$}

In this subsection we obtain $so(\ell+1,\ell+d)$-analogues of the vector generators $C_i^{(n)}$.

\vskip 0.2cm

Perhaps, the simplest way to obtain counterparts of $C_i^{(n)}$ consists of two steps. First, we find an analogue for $n=\ell$. According to \eqref{Galilei}, $C_i^{(\ell)}$ transforms as a vector under rotations and has zero weight. There is only one basis element which has these properties:
\bea
&&
C_i^{(\ell)} \equiv \mathbb{M}_{\ell,2\ell+i}, \quad i = \overline{1,d}.
\eea

At the next step, we obtain other vector generators in the following way
\bea\label{C}
&&
\begin{aligned}
&
C_i^{(\ell\pm k)} = (\mp1)^{k}\frac{(\ell-k)!}{\ell!}\underbrace{[L_{\pm 1},[L_{\pm 1},...[L_{\pm 1},C_i^{(\ell)}]...]]}_{\mbox{$k$ times}} =
\\[2pt]
&
= \mp\frac{(-2)^{k-1}(\ell-k)!}{\ell!} \left(\prod_{\gamma=1}^{k}\eta_{\ell-\gamma}^{\pm}\right)(\mathbb{M}_{2\ell-k,2\ell+i} \mp \mathbb{M}_{k,2\ell+i}),
\end{aligned}
\eea
where $k = \overline{1,\ell}$.

It can be directly verified that the commutators between these generators and $M_{ij}$ in \eqref{M}, $L_0$ in \eqref{L0}, $L_{\pm1}$ in \eqref{Lpm1} are the same as in \eqref{Galilei}.  But here the vector generators \eqref{C} do not form an abelian ideal. Note that the generators \eqref{C} form a new basis in the subspace of $so(\ell+1,\ell+d)$ spanned by $\mathbb{M}_{\gamma,2\ell+i}$ with $\gamma=\overline{0,2\ell}$, $i=\overline{1,d}$.

\section{The In\"{o}n\"{u}-Wigner contraction of $so(\ell+1,\ell+d)$}\label{sect4}

In the previous section we have constructed a new basis of the conformal algebra $so(\ell+1,\ell+d)$. Here we use this basis to formulate the contraction procedure for $so(\ell+1,\ell+d)$ which results in the $\ell$-conformal Galilei algebra as a subalgebra of a contracted algebra. The case $\ell=2$ is treated in more detail. In particular, we show how the realization \eqref{so_gen} of $so(3,2+d)$ can be contracted.

\vskip 0.2cm

In general, nonrelativistic conformal algebras are not semisimple \cite{Henkel,Negro_1}. Their semisimple part is given by $so(1,2)\oplus so(d)$. For this reason, let us introduce the contraction procedure of the algebra $so(\ell+1,\ell+d)$ with respect to its subalgebra spanned by $L_{-\pm1}$, $L_{0}$, and $M_{ij}$. To this end, we redefine other generators as follows:
\bea\label{redef1}
&&
C_i^{(n)} \to \frac{1}{c}C_i^{(n)}, \qquad T_{l}^{(n)} \to \frac{1}{c}T_l^{(n)},
\eea
where $c$ is a constant which may be interpreted as the speed of light. It is obvious that the limit $c\to\infty$ in the structure relations of $so(\ell+1,\ell+d)$ yields the algebra whose nonvanishing commutators involve \eqref{Galilei} and \eqref{LT}. So, the $\ell$-conformal Galilei algebra in $d$ spatial dimensions as well as the $l$-conformal Galilei algebras in one spatial dimension spanned by $\{L_{0},L_{\pm1},T_{l}^{(n)}\}$ with $l=3$, $5$, ..., $(2\ell-1)$ appear in the contracted algebra as subalgebras\footnote{It should be noted that the so-called $\ell$-conformal Newton-Hooke algebra \cite{Negro_1} for integer $\ell>1$ may be derived from $so(\ell+1,\ell+d)$ in the same manner. For this, we have to redefine $L_{-1}\to L_{-1}+\Lambda L_{1}$ in addition to \eqref{redef1}, where $\Lambda$ is a nonrelativistic cosmological constant. The $\mathbb{IW}$ contraction of the conformal algebra which leads to the $\ell=1$-conformal Newton-Hooke algebra was introduced in Ref. \cite{AV_2011}.}.

Let us consider the case $\ell=2$ in more detail and show how we may obtain a realization of the contracted algebra from \eqref{so_gen} by applying a contraction. To this end, we write down the following representation of the generator $L_0$:
\bea\label{repL0}
&&
L_{0} = 2D + J_{02}.
\eea
With the aid of the analysis in Sect. \ref{sect3}, we obtain:
\bea\label{repLpm1}
&&
L_{-1} = \eta_0^{-}(P_{0} - P_{2}) + \eta_1^{-}(J_{01} - J_{12}), \quad L_{1} = \frac{1}{\eta_0^{-}}(K_{0} + K_{2}) + \frac{3}{\eta_1^{-}}(J_{01} + J_{12}).
\eea
It can be verified straightforwardly that the change of coordinates
\bea
&&
y^{0} = \eta_0^{-}\left((\eta_1^{-})^{2}s_2+s_0\right), \quad y^{1} = \eta_0^{-}\eta_{1}^{-}s_1, \quad y^{2} = \eta_{0}^{-} \left((\eta_1^{-})^{2}s_2-s_0\right), \quad y^{2+i} = \eta_0^{-}\eta_{1}^{-} x_i, \quad i = \overline{1,d},
\nonumber
\eea
absorbs the constants $\eta_{0}^{-}$, $\eta_{1}^{-}$ in \eqref{repLpm1}. In terms of new coordinates the expressions \eqref{repL0}, \eqref{repLpm1} take the form
\bea\label{Ls}
&&
\begin{aligned}
&
L_{-1} = \frac{\partial}{\partial s_{0}} + 2s_0 \frac{\partial}{\partial s_1} - s_1 \frac{\partial}{\partial s_2},\quad 
L_{0} = s_0 \frac{\partial}{\partial s_0} + 2s_1 \frac{\partial}{\partial s_1} + 3 s_2 \frac{\partial}{\partial s_2} + 2 x_i \frac{\partial}{\partial x_i},
\\[2pt]
&
L_{1} = (x_i x_i - s_1^2)\frac{\partial}{\partial s_2} + 4s_0 \left(s_0 \frac{\partial}{\partial s_0} + s_1 \frac{\partial}{\partial s_1} + x_i \frac{\partial}{\partial x_i}\right) - 3s_1\frac{\partial}{\partial s_0} + 6 s_2 \frac{\partial}{\partial s_1}.
\end{aligned}
\eea

The desire to have the standard form of the generator of time translations $L_{-1}$ motivates us to apply the straightening out theorem for this vector field. As a result, we obtain the following representation of the generators $L_0$, $L_{\pm1}$:
\bea\label{Lt}
&&
\begin{aligned}
&
L_{-1} = {\color{blue}\frac{\partial}{\partial t}}, \quad L_0 = {\color{blue}t \frac{\partial}{\partial t} + 2 r_1 \frac{\partial}{\partial r_1} + 3 r_2\frac{\partial}{\partial r_2} + 2 x_i\frac{\partial}{\partial x_i}},
\\[2pt]
&
L_{1} = {\color{blue}-t^{2}L_{-1} + 2 t L_{0}} + {\color{red} 6 r_2 \frac{\partial}{\partial r_1}} - 3 r_1 \frac{\partial}{\partial t}  + (x_i x_i - 4 r_1^2)\frac{\partial}{\partial r_2},
\end{aligned}
\eea
where the coordinates $t$, $r_1$, and $r_2$ are defined by\footnote{When verifying \eqref{Ls} and \eqref{Lt}, the identities
\bea
&&
\begin{aligned}
&
\frac{\partial}{\partial y^{\bar{0}}} = \frac{1}{2\eta_0^{-}}\left(\frac{1}{(\eta_{1}^{-})^{2}}\frac{\partial}{\partial s_{2}} + \frac{\partial}{\partial s_{0}}\right), && \frac{\partial}{\partial y^{\bar{1}}} = \frac{1}{\eta_0^{-}\eta_{1}^{-}}\frac{\partial}{\partial s_1}, && \frac{\partial}{\partial y^{\bar{2}}} = \frac{1}{2\eta_0^{-}}\left(\frac{1}{(\eta_1^{-})^{2}}\frac{\partial}{\partial s_{2}} - \frac{\partial}{\partial s_{0}}\right),
\\[2pt]
&
\frac{\partial}{\partial s_0} = \frac{\partial}{\partial t} - 2 t \frac{\partial}{\partial r_1} + r_1 \frac{\partial}{\partial r_2} - t^2 \frac{\partial}{\partial r_2}, && \frac{\partial}{\partial s_1} = \frac{\partial}{\partial r_1} + t\frac{\partial}{\partial r_2}, && \frac{\partial}{\partial s_2} = \frac{\partial}{\partial r_2},
\end{aligned}
\nonumber
\eea
prove to be helpful.}
\bea
&&
t = s_0, \quad r_1 = s_1 - s_0^2, \quad r_2 = s_2 + s_0 s_1 - \frac{2s_0^3}{3}.
\nonumber
\eea

Based on the results in Sect. \ref{sect3}, we obtain the following operators corresponding to $T_{3}^{(n)}$, $C_i^{(n)}$, and $M_{ij}$:
\bea\label{Ts}
&&
\begin{aligned}
&
T_{3}^{(0)} = \eta_0^{-} (\eta_1^{-})^{2} (P_{0} + P_{2}) = {\color{blue} \frac{\partial}{\partial r_2}}, \;
T_{3}^{(1)} = \eta_0^{-} \eta_{1}^{-} P_{1} =  {\color{blue} t\frac{\partial}{\partial r_2}} + {\color{red} \frac{\partial}{\partial r_1}},
\\[2pt]
&
T_{3}^{(2)} = -\frac{3\eta_0^{-}}{5}(P_{0}-P_{2}) + \frac{2\eta_{1}^{-}}{5} (J_{01}-J_{12}) =  {\color{blue}t^2\frac{\partial}{\partial r_2}} + {\color{red}2t\frac{\partial}{\partial r_1}} - r_1\frac{\partial}{\partial r_2} - \frac{3}{5}\frac{\partial}{\partial t},
\\[2pt]
&
T_{3}^{(3)} = -\frac{3}{5}(D-2J_{02})
= {\color{blue}t^3 \frac{\partial}{\partial r_2}} + {\color{red}3 t^2\frac{\partial}{\partial r_1}} - 3t r_1 \frac{\partial}{\partial r_2} - \frac{3}{5}\left(3 t\frac{\partial}{\partial t} + r_1\frac{\partial}{\partial r_1} - r_{2}\frac{\partial}{\partial r_2} + x_i\frac{\partial}{\partial x_i}\right),
\\[2pt]
&
T_{3}^{(4)} = -\frac{3}{5\eta_0^{-}}(K_{0} + K_{2}) + \frac{6}{5\eta_{1}^{-}} (J_{01} + J_{12}) = {\color{blue}t^{4}\frac{\partial}{\partial r_2}} + {\color{red}4 t^3\frac{\partial}{\partial r_1}} - 6 t^2 r_1 \frac{\partial}{\partial r_2} - \frac{6}{5}(r_1 + 3t^2)\frac{\partial}{\partial t} -
\\[2pt]
&
- \frac{12}{5}(tr_1 - r_2)\frac{\partial}{\partial r_1} - \frac{3}{5}(r_1^2 - 4tr_2 + x_i x_i) \frac{\partial}{\partial r_2} - \frac{12}{5} tx_i \frac{\partial}{\partial x_i},
\\[2pt]
&
T_{3}^{(5)} = -\frac{3}{\eta_0^{-}\eta_1^{-}}K_{1} = {\color{blue}t^{5}\frac{\partial}{\partial r_2}} + {\color{red}5 t^{4}\frac{\partial}{\partial r_1}} - 6t(r_1 + t^2) \frac{\partial}{\partial t} - 3 (t^2 r_1 - 4tr_2 + r_1^2 + x_i x_i) \frac{\partial}{\partial r_1} -
\\[2pt]
&
- (10t^3 r_1 - 6t^2 r_2 + 3t r_1^2 + 6 r_1 r_2 + 3t x_i x_i) \frac{\partial}{\partial r_2} - 6 (r_1 + t^2) x_i \frac{\partial}{\partial x_i},
\\[2pt]
&
T_{3}^{(6)} = \frac{9}{\eta_0^{-}(\eta_1^{-})^{2}} (K_{0} - K_{2}) = {\color{blue}t^{6} \frac{\partial}{\partial r_2}} + {\color{red}6t^5 \frac{\partial}{\partial r_1}} + 9(x_i x_i - (r_1 + t^2)^{2}) \frac{\partial}{\partial t} - 
\\[2pt]
&
- 12 (t^3 + 3tr_1 - 3r_2) x_i \frac{\partial}{\partial x_i} - 6(2t^3 r_1 - 6 r_2 (t^2 - r_1) + 3t (x_i x_i + r_1^2)) \frac{\partial}{\partial r_1} -
\\[2pt]
&
- 3(5t^4 r_1 - 4t r_2 (t^2 - 3r_1) + 3 r_1^2 (t^2 + r_1) + 3 x_i x_i (t^2 - r_1)   - 12 r_2^2)\frac{\partial}{\partial r_2},
\\[2pt]
&
C_i^{(0)} = \eta_0^{-} \eta_{1}^{-} P_{2+i} = {\color{blue}\frac{\partial}{\partial x_i}}, \quad C_i^{(1)} = \frac{\eta_{1}^{-}}{2}(J_{0,2+i}+J_{2,2+i}) = {\color{blue}t\frac{\partial}{\partial x_i}} + \frac{1}{2}x_i\frac{\partial}{\partial r_2},
\\[2pt]
&
C_i^{(2)} = J_{1,2+i} = {\color{blue}t^2 \frac{\partial}{\partial x_i}} + r_1 \frac{\partial}{\partial x_i} + x_i \left(\frac{\partial}{\partial r_1} + t \frac{\partial}{\partial r_2}\right),
\\[2pt]
&
C_i^{(3)} = -\frac{3}{2\eta_{1}^{-}} (J_{0,2+i}-J_{2,2+i}) = {\color{blue}t^{3}\frac{\partial}{\partial x_i}} + (3t r_{1} - 3r_2) \frac{\partial}{\partial x_i} - \frac{3}{2} x_i \left(\frac{\partial}{\partial t} - 2t\frac{\partial}{\partial r_1} +r_1 \frac{\partial}{\partial r_2} - t^2 \frac{\partial}{\partial r_2}\right),
\\[2pt]
&
C_i^{(4)} = \frac{3}{\eta_0^{-}\eta_{0}^{-}} K_{2+i} = {\color{blue}t^{4}\frac{\partial}{\partial x_i}} - (3r_1(r_1 - 2t^2) + 12t r_2 - 3x_j x_j) \frac{\partial}{\partial x_i} + 
\\[2pt]
&
+ 2x_i \left((t^3-3tr_1 - 3r_2)\frac{\partial}{\partial r_2} - 3t\frac{\partial}{\partial t} + 3(t^2-r_1)\frac{\partial}{\partial r_1} - 3 x_j \frac{\partial}{\partial x_j}\right),
\\[2pt]
&
M_{ij} = -J_{2+i,2+j} = {\color{blue}x_i\frac{\partial}{\partial x_j} - x_j\frac{\partial}{\partial x_i}}.
\end{aligned}
\eea

Let us apply the rescaling $t\to ct$ and redefine the generators as follows
\bea\label{redef2}
&&
c^{-n} L_{n} \to L_{n}, \qquad c^{-m}T_{3}^{(m)} \to T_{3}^{(m)}, \qquad c^{-m} C_i^{(m)} \to C_i^{(m)}.
\eea
The only blue terms in the RHS of \eqref{Ls} and \eqref{Ts} do not vanish in the limit $c\to\infty$. Taking into account the realization \eqref{Galilei_gen}, it is evident that the generators obtained in such a way form the contracted algebra \eqref{Galilei}, \eqref{LT} for $\ell=2$.

It is interesting to note that we may use the rescaling of another type: $t\to ct$, $r_1\to c^{-1}r_1$. Then after redefinition \eqref{redef2}, the limit $c\to\infty$ leaves the blue and red terms in RHS of \eqref{Lt}, \eqref{Ts} unchanged, while others vanish. The nonzero commutators of derived generators consist of the same relations \eqref{Galilei}, \eqref{LT} for $\ell=2$.

As we can see, the nonrelativistic spacetime, where the contracted algebra is realized, is parameterized by the temporal $t$, spatial $x_i$ and additional $r_1$, $r_2$ coordinates. We expect that the meaning of the latter degrees of freedom may be clarified by investigating the nonrelativistic limit of $so(3,2+d)$-invariant systems. This subject is left out of the present paper.

\section{Concluding remarks}\label{sect5}

In this work we have shown that the $\mathbb{IW}$ contraction of $so(\ell+1,\ell+d)$ for the integer $\ell>1$ may lead to an algebra which contains the variety of conformal extensions of the Galilei algebra as subalgebras. These extensions involve the $\ell$-conformal Galilei algebra in $d$ spatial dimensions as well as the $l$-conformal Galilei algebras in one spatial dimension with $l=3$, $5$, ..., $(2\ell-1)$. This was achieved by constructing the appropriate basis of $so(\ell+1,\ell+d)$.

The $so(\ell+1,\ell+d)$-analogue of the generator of dilatations $L_0$ \eqref{L0} has played the central role in our considerations. The adjoint action of this generator has allowed us to associate each element of $so(\ell+1,\ell+d)$ with its weight. In fact, by using this correspondence we have obtained a grading of $so(\ell+1,\ell+d)$. The grading has enabled us to introduce the ansatz \eqref{Lpm1} for $so(\ell+1,\ell+d)$-analogues of $L_{\pm1}$ and find them. In turn, the generators $L_{-1}$ and $L_{1}$ have acted as the weight lowering and raising operators respectively. When considering the subalgebra $so(\ell+1,\ell)$, $L_{\pm1}$ has helped to reveal the $l$-conformal Galilean structures with $l=3$, $5$, ..., $(2\ell-1)$ \eqref{LT}. Besides that, we have obtained $so(\ell+1,\ell+d)$-counterparts for the chain of the vector generators $C_i^{(n)}$ \eqref{C} with the aid of the ladder operators.

Considering the realization of $so(3,d+2)$ in flat spacetime, we have demonstrated that free parameters $\eta_0^{-}$, $\eta_1^{-}$ which appeared when constructing $so(3,d+2)$-analogues of $L_{\pm1}$ \eqref{Lpm1}, can be absorbed by applying an appropriate change of coordinates. We have also proposed two contraction schemes for the realization \eqref{so_gen} of the conformal algebra $so(3,d+2)$ in flat spacetime.

Note that the same algebra, whose nonvanishing relations consist of \eqref{Galilei} and \eqref{LT}, may be obtained from the $so(\ell,\ell+d+1)$ via $\mathbb{IW}$ contraction by applying the steps presented in Sect. 3 and Sect. 4. Indeed, the following change of the metric \eqref{eta}
\bea
&&
\eta_{\alpha\beta}= \mbox{ diag} (\underbrace{-,-,..,-}_{\mbox{$\ell$ times}},\underbrace{+,+,..,+}_{\mbox{$\ell+d+1$ times}})
\nonumber
\eea
does not affect almost any of the results obtained in subsection \ref{subs3.1}. Only commutators between $\Omega_{\alpha}^{\pm}$ in \eqref{comrel} are slightly altered as follows:
\bea
&&
[\Omega_{\alpha}^{\pm},\Omega_{\beta}^{\pm}] = \Sigma_{\alpha\beta}^{\pm}, \qquad [\Omega_{\alpha}^{\pm},\Omega_{\beta}^{\mp}] = \Lambda_{\alpha\beta}.
\nonumber
\eea
This entails minor changes in \eqref{sys}, \eqref{sol}, \eqref{sys2}, and \eqref{C} which, however, lead to the same contracted algebra. In particular, taking into account the analysis in Sect. \ref{sect4}, we conclude that the $l=3$ conformal Galilei algebra in one spatial dimension (spanned by $L_n$ and $T_{3}^{(m)}$ with $n=0,\pm1$, $m=\overline{0,6}$) can be obtained from the anti-de Sitter algebra $so(2,3)$ via a contraction.

During the last fifteen years or more, interest in the $\ell=1$ conformal Galilei algebra was largely inspired by the fact that this algebra may appear as the nonrelativistic counterpart of $so(2,1+d)$. In particular, the conformal Galilei algebra was widely used within the context of the nonrelativistic AdS/CFT-correspondence \cite{Bagchi}-\cite{Martelli} where the contraction procedure plays important role. It can be expected that the results in the present work may help to draw attention to the $\ell$-conformal Galilei algebra with the integer $\ell>1$ in connection with the study of the nonrelativistic holography.

As was mentioned in the introduction, we have analysed the algebra $so(\ell+1,\ell+d)$ due to the observation which was made in the work \cite{Negro_2}. But of course, it is worth probing other candidates to consider relativistic counterpart of the $\ell$-conformal Galilei algebra. To find them it looks promising to use the approach developed in the work \cite{Jose}, where conformal symmetries are investigated via deformation theory.

\fontsize{10}{13}\selectfont

\end{document}